# Uncertainty and Exploration of Deep Learning-based Atomistic Models for Screening Molten Salt Properties and Compositions

Author Name(s): Stephen T. Lam, Shubhojit Banerjee, Rajni Chahal
*University of Massachusetts Lowell*
*1 University Ave, Lowell, MA, 01854, USA*
*Tel: 978-934-3150, Email: Stephen_Lam@uml.edu*

**Abstract –** *Due to extreme chemical, thermal, and radiation environments, existing molten salt property databases lack the necessary experimental thermal properties of reactor-relevant salt compositions. Meanwhile, simulating these properties directly is typically either computationally expensive or inaccurate. In recent years, deep learning (DL)-based atomistic simulations have emerged as a method for achieving both efficiency and accuracy. However, there remain significant challenges in assessing model reliability in DL models when simulating properties and screening new systems. In this work, structurally complex LiF-NaF-ZrF$_4$ salt is studied. We show that neural network (NN) uncertainty can be quantified using ensemble learning to provide a 95% confidence interval (CI) for NN-based predictions. We show that DL models can successfully extrapolate to new compositions, temperatures, and timescales, but fail for significant changes in density, which is captured by ensemble-based uncertainty predictions. This enables improved confidence in utilizing simulated data for realistic reactor conditions, and guidelines for training deployable DL models.*

## I. INTRODUCTION

Molten salt reactors offer substantial improvements in safety and economics over existing systems. However, due to difficulties in collecting experimental data in extreme (high temperature, irradiation, corrosion) environments, the existing molten salt properties databases lack adequate transport and thermophysical properties at reactor-relevant salt compositions [1]. Accurately simulating the structure and properties of these systems is challenging due to computational scaling ($O(N^3)$) limitations in electronic structure methods based on density functional theory (DFT) and cost-accuracy tradeoffs of classical interatomic potential methods. Here, ab initio-accurate deep learning (DL) interatomic potentials that scale linearly with system size can circumvent these limitations. However, model reliability remains a challenge in utilizing DL models under new conditions outside of the training set.

As such, there is a strong motivation for developing methods that accurately characterize the uncertainty of neural networks (NN) for atomistic modeling. In DL applications, popular methods include Bayesian inference, and ensemble learning [2]. While a Bayesian approach provides a clear description of parametric uncertainty in the model, high computational effort and memory would add significant overhead to atomistic simulations, which require millions of NN forward pass calculations to simulate properties. Meanwhile, ensemble methods are readily implemented and parallelizable, making them suitable for molecular dynamics (MD). However, their ability to predict true error in molten salts has yet to be studied.

## II. MATERIALS AND METHODS

DFT calculations were performed using Vienna Ab-Initio Simulation Package (VASP), with the projector augmented wave (PAW) method, plane wave basis set with 650 eV cutoff, and gamma point-only sampling, based on previous studies [3]. The deep learning potential was trained using DeepMD-kit [4]. The training set for the DL potential consisted of 28,261 DFT calculations of molten 42-29-29 mol% LiF-NaF-ZrF$_4$ and 26-37-37 LiF-NaF-ZrF$_4$ sampled at ~1000K under the canonical ensemble [3]. Each supercell in the training set contained 84-89 atoms. NN-based MD simulations were performed with large supercells of up to 43 Å, and 2 nanoseconds to demonstrate the ability of the network to extrapolate to larger systems and longer timescales. For the test set, configurations were sampled far outside of the training set with 38-51-11 mol% LiF-NaF-ZrF$_4$, T=2000K, and expanded supercells by up to 120%.

For uncertainty quantification, an ensemble of neural networks were trained simultaneously with different randomized parameter initializations. The model deviation relative to DFT is taken as the raw uncertainty ($\mu_i$) for a given configuration $i$, and compared against the true error ($y_i$). Different scaling methods and functions can be used to calibrate the model uncertainty against the true error. Here, the raw uncertainties are scaled by a linear function that minimizes the sum of squared errors $\Sigma_n(y_i - \mu_i)$ for configurations $n$ exhibiting the top 5% error. This produces a conservative estimation (~95% CI) of potential error, providing an efficient method for error-checking during MLMD simulations.

## III. RESULTS AND DISCUSSION

The trained NN-based calculations reproduced DFT calculations for training set configurations with energy error of less than 3 meV/atom, and force error averaging 70 meV/Å. Similar errors are seen for test set configuration when only the composition and temperature are changed. A structural comparison between the training and test set is shown in Fig. 1.

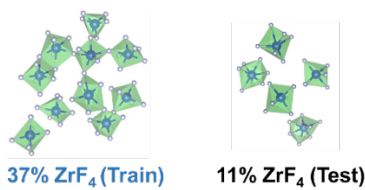

Fig. 1. Train versus test configurations at 1000K. The ZrF polyhedra are shown with Li and Na hidden.

While training compositions with high $ZrF_4$ content show significant network effects, salts with only 11% $ZrF_4$ exhibit isolated $ZrF_4$ monomers. Moreover, average fluorine coordination increases from 6.5 for 37% $ZrF_4$ to 7.0 11% $ZrF_4$ which is supported by Raman spectra observations across composition and temperature [3]. As such, the DL model accurately captures the diverse coordination chemistries while demonstrating robustness against compositional change. Here, the uncertainty metric is validated as it remains low under these conditions as shown in datasets A (Train), and B/C (Test) of Fig. 2.

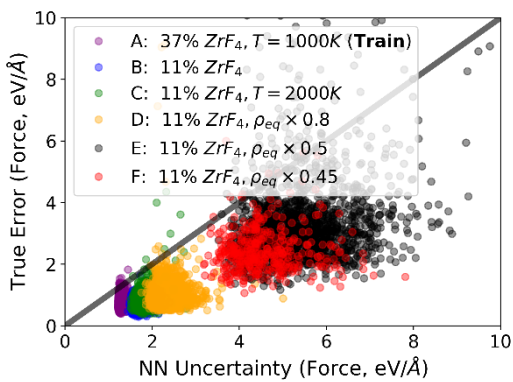

Fig. 2. True error vs. scaled NN uncertainty for training and test configurations

A high error (order of magnitude) is observed when simulation cells are subject to volumetric expansion of up to 120% as shown in datasets D/E/F (Test) of Fig. 2. This demonstrates that while NN model can extrapolate across extreme changes in composition and temperature, it can fail to extrapolate across different pressure and densities.

Here, we demonstrate that the scaled uncertainty successfully provides a specified upper bound on the true error (in this case 95%). Moreover, we show the uncertainty method is sensitive to errors in equation of state (P,V,T). This suggests that it will perform well for estimating uncertainty of simulated thermodynamic properties. As such, further investigation is currently underway for various properties and salt mixtures (e.g., other prototypical salts, actinides, etc.).

Lastly, it is shown that the model can be used reliably when the predicted uncertainty is low, reducing the need for constant validation against computationally expensive DFT calculations. To reduce overall uncertainty, expanded cells in the training set are recommended when developing DL interatomic models. In other cases where model uncertainty is high, the model can be systematically improved by adding more training data in the high-uncertainty regions.

## IV. CONCLUSIONS

Deep learning potentials and uncertainty models were developed for prototypical molten salt LiF-NaF-$ZrF_4$. The potentials accurately extrapolated across composition and temperature while exhibiting a high error across densities, highlighting the need for increased sampling of variable cell volumes for developing robust potentials. Using ensemble learning, uncertainty can be quantified and scaled to provide accurate bounds of the true error. This could enable deployment and integration of NN models for calculating properties used in MSR reactor design and optimization.

## ACKNOWLEDGMENTS

This work is supported by DOE-NE's Nuclear Energy University Program (NEUP) under Award DE-NE0009204, and resources provided by the National Energy Research Scientific Computing Center under awards ASCR-ERCAP0022362 and BES-ERCAP0022445.